# Glass-based charged particle detector performance for Horizon-T EAS detector system

A. Duspayev [1a], M. Yessenov [1a], D. Beznosko [2a], A. Iakovlev [a], M.I. Vildanova [b], V.V. Zhukov [b]

(a) Nazarbayev University, Astana, Kazakhstan.

(b) P. N. Lebedev Physical Institute of the Russian Academy of Sciences, Moscow, Russia.

## Abstract

An implementation of a novel of glass-based detector with fast response and wide detection range is needed to increase resolution for ultra-high energy cosmic rays detection. Such detector has been designed and built for the Horizon-T detector system at Tien Shan high-altitude Science Station. The main characteristics, such as design, duration of the detector pulse and calibration of a single particle response are discussed.

## 1. Introduction

"Horizon-T" detector system is constructed to study Extensive Air Shower (EAS) caused by cosmic particles of energies higher than $10^{16}$eV [1]. The system consists of eight charged particle detection points and one Vavilov-Cherenkov detector located at Tien Shan high-altitude Science Station, a part of P.N. Lebedev Physical Institute of the Russian Academy of Sciences. "Horizon-T" is used to study space-time distribution of the charged particles in EAS disk, and Vavilov-Cherenkov radiation from it. Each detection point consists of scintillator detector (SD) constructed from the 1 m$^2$ plastic scintillator [2] that is 5 cm thick and is read out by Hamamatsu [3] PMT R7723, and glass detector (GD) with thick optical glass [4] that is further discussed in this paper. They register time distribution of the charged particles density during EAS disk passage through each detection point. Data is analyzed by the novel method using timing information from signal pulse shape from each detector.

## 2. Cherenkov glass detector physics in simulation

Glass is a widely used as a medium for high-energy physics (HEP) experiments for Cherenkov detector construction. This section briefly revisits physics behind Cherenkov detectors from the point of view of construction a simulation.

As particle travels through the glass, total number of emitted photons is calculated using standard Bethe-Bloch formula in a simplified form [5]:

$$\frac{dN}{dx} = 2\pi\alpha \left(\frac{1}{\lambda_2} - \frac{1}{\lambda_1}\right)\left(1 - \frac{1}{\beta^2 n^2}\right) \qquad (1)$$

---

[1] Equal contribution from both authors
[2] dima@dozory.us

where $\frac{dN}{dx}$ - number of particles per unit length, $\alpha$ - fine structure constant, $\lambda_1, \lambda_2$ - wavelengths corresponding to the maximum energy range of the most efficient performance of PMT ($\lambda_1 \approx 300nm, \lambda_2 \approx 500nm$), $\beta \approx 1$ - corresponding velocity of the incoming particles, $n \approx 1.6$ - glass refraction index. In principle, refraction index is a function of wavelength, but it can be averaged over a small wavelengths range for simplicity as the small differences don't impact the simulation outcome.

Photon emission along progenitor particle's trajectory can be considered as a Poisson process with mean equal to the product of the average number of emitted photons per mm (~40 photons per mm from eq. 1) and the step of propagation. When applied for a 3cm path length, calculations result in ~1000-1500 photons being emitted for each particle. Each photon is assigned a uniform polar angle $\varphi$ and angle $\theta = \cos^{-1}\frac{1}{\beta n}$ (known as Cherenkov angle, $\approx 51.3^0$ for $n \approx 1.6$) and with respect to the trajectory of the progenitor that are converted into coordinate system associated with glass. Due to glass thickness, the initial photon emission time is same as the parent particle passage time ~ 100 ps.

As light reaches the sides of the glass, depending on the incident angle and whether side is painted or not different processes may occur. For unpainted sides, photon is reflected back if its angle of incidence is larger than the critical angle, or escapes the glass at the refracted angle obtained from Snell's law. Critical angle value is determined by the total internal reflection condition: $sin\,\theta_{crit} = \frac{n_{air}}{n_{glass}}$. For $n_{glass} \approx 1.6$ and $n_{air} \approx 1$ we get $\theta_{crit} \approx 38.7^0$. For the painted sides diffusive reflection modeling is used for obtaining new photon zenith and polar angles [6]. Photons can also be absorbed at the detector sides and it is implemented as a fixed absorption probability.

## 3. Simulation of the GD

### 3.1 Detector module description

The simulation of the detector module was done using ROOT framework [7]. The simulation code has been initially developed in order to determine the most appropriate materials and geometry for the detector module construction for HorizonT-Kazakhstan (HT-KZ) experiment [8], a distributed detector system under R&D at Nazarbayev University (NU), Astana, Kazakhstan. The same simulation code has been used to check the properties of the developed GD design and to compare results of the simulation with the calibration data obtained for the Horizon-T experiment [9], specifically the light detection uniformity from different parts of detection volume. The graphical representation of the simulated glass detector is shown in Figure 1.

The sides of the glass are painted black to reduce multiple reflections, whereas the bottom side of the glass is painted white to reflect photons diffusively. Such model results in more efficient light detection and has already been used for HT-KZ detector modules simulation [8]. Diffusion reflection is used for reflection from the painted sides, modeling the angle of reflection discussed above.

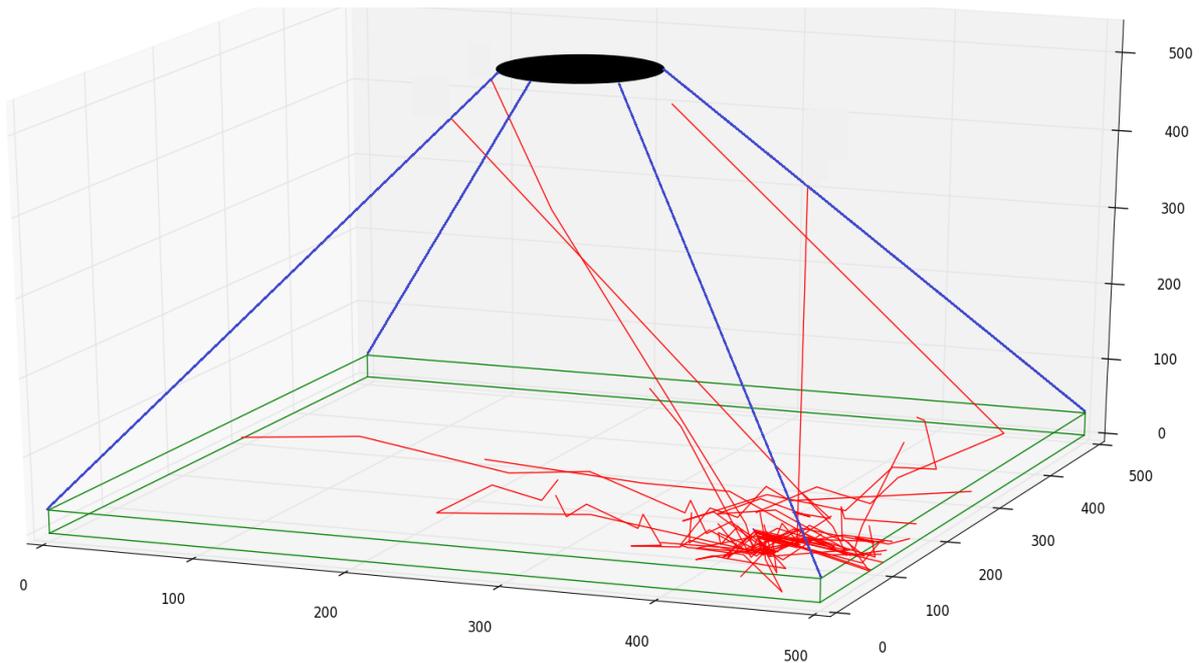

**Figure 1: Glass detector schematics: 3 cm thick glass with 0.5m square base (sides are shown green), PMT (~20% efficiency, ~5cm radius) located above the base (black disk at the top) and 0.5m height shell between two components (sides are shown blue).**

## 3.2 Simulation algorithm

A sample of ~$10^5$ particles is processed for each run, where each particle gets assigned linearly distributed random initial *(x,y,z)* - coordinates on the top face of the glass base and uniform zenith (within selected range) and polar angles with respect to the glass surface for the total of five random variables. Then every photon that is produced along the particle track is propagated till it reaches the side of the glass and probabilities of it being reflected, absorbed or escaped from the glass are determined using Monte-Carlo process [10]. Only photons that exit the glass from the PMT side are propagated towards PMT region: they can be absorbed at the sides of the shell along the way or reach PMT region and saved as 'detected'. The size of the propagation step can be varied (0.5mm for the results presented); the influence of the step size on the result is taken as a systematic error. Two histograms are produced: number of detected photons vs initial *(x,y)* - coordinates of their parent particle and time of photons arrival to PMT for each detected photon. The first histogram indicates the spatial probability of photons reaching the PMT from different parts of the glass; the second is used to determine the width of light pulse detected by PMT (no internal PMT effects are applied). These results are used to qualitatively assess performance of various module arrangements and the detected signal width.

## 3.3. Simulation validity

Simulation of SD has been tested to check the simulation validity and its correspondence to the experimental measurements. A sample of $5*10^4$ photons has been used. The obtained spatial distribution of the detected photons is in Figure 2:

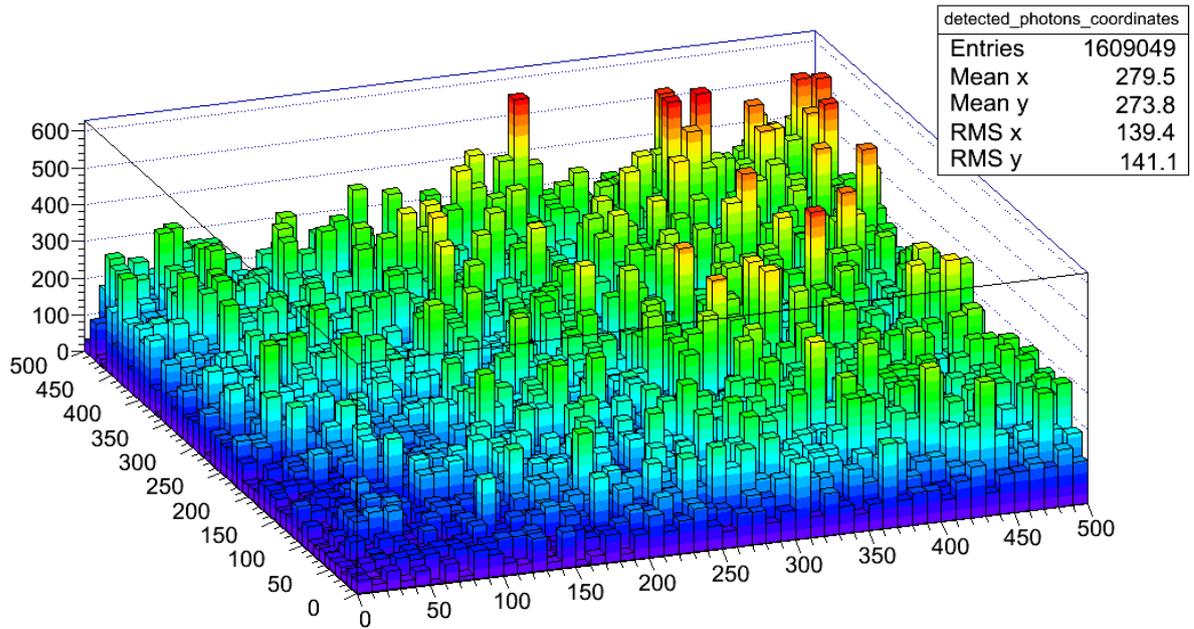

**Figure 2: Spatial distribution of the detected photons vs initial coordinates of the progenitor particle for SD.**

A light yield of ~ 32 detected photons per particle has been obtained. This simulation data is used to qualitatively assess the non-uniformity of light detection by the PMT in this geometry. First, we divide histogram of spatial distribution of detected photons by histogram of spatial distribution of progenitor particles. Then, we take number of detected photons per particle along 8 separate lines: both diagonals, x = 150, 250, 350, y = 150, 250, 350mm. We get 8 1D-histograms, all of them are normalized such that maximum of each histogram is 1. Then, the mean and the standard deviation for each histogram are calculated. Uniformity coefficient (average weight) is computed as a mean of all 8 means calculated previously. For SD it is equal to 0.69±0.11, which corresponds to the experimental measurement mentioned in section 4.2.

### 3.4. Simulation results

#### 3.4.1. PMT placement

At first stage, option for PMT placement has been tested: first - with PMT above glass (Figure 3), second - with PMT below glass (Figure 4). A sample of $10^5$ particles has been used for both cases. Simulation results show that the first option gives more efficient light yield (≈3.65 detected photons per particle) than the option with PMT below glass does (≈3.19 detected photons per particle), thus, the first geometry has been used for the next stages.

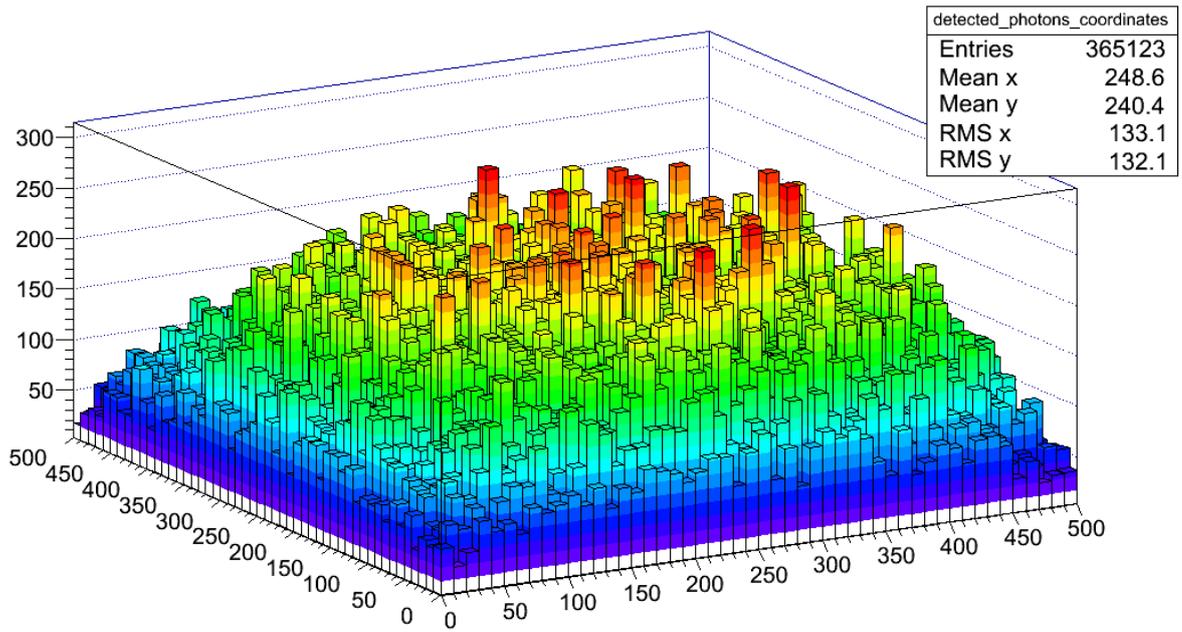

**Figure 3: Spatial distribution of the detected photons vs initial coordinates of the progenitor particle for GD with PMT above.**

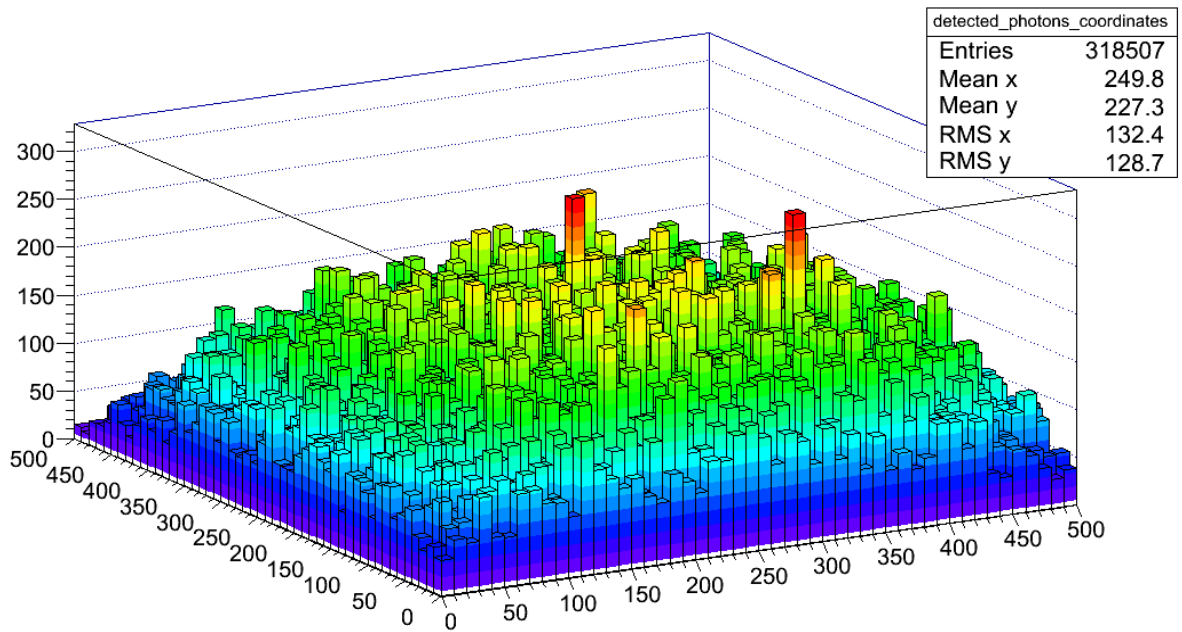

**Figure 4: Spatial distribution of the detected photons vs initial coordinates of the progenitor particle for GD with PMT below.**

### 3.4.2. Detector response uniformity

From Figure 3, a total of ~$3.65*10^5$ photons are detected. More photons are detected from the particles passing near the glass center. One can notice that the peak of the distribution is displaced from the geometric center. This is due to the Cherenkov angle, because the emitted photons are not concentrated at one place, but they are spread over corresponding area. Notably less detected photons originate from the particles passing near corners. Such behavior is expected, since the sides of the glass are painted black to reduce the pulse width. For chosen arrangement of GD the average weight of 0.77±0.02 has been obtained. This value can be used for calibration of GD minimum ionizing particle response.

### 3.4.3. Pulse width

The distribution of the PMT detected photons arrival time is shown in Figure 5. Full width at half-maximum of the distribution is ~ 1.5ns. This value is much bigger than the light production time in glass (~ 100ps). Such increase can be explained by the fact that the detected photons experience many reflections on the sides of the glass and the shell and require additional time to arrive to the PMT. The distribution can be fitted by the following function:

$$f(t) = \int_0^\infty p(t')g(t-t')dt', \qquad (2)$$

where $p(t')$ - photon propagation function, which depends on atomic properties of the detector's materials, $g(t) = \frac{1}{\sqrt{2\pi}\sigma} e^{-\frac{t^2}{2\sigma^2}}$. Detailed derivation and discussion of eq. 2 can be found in [11].

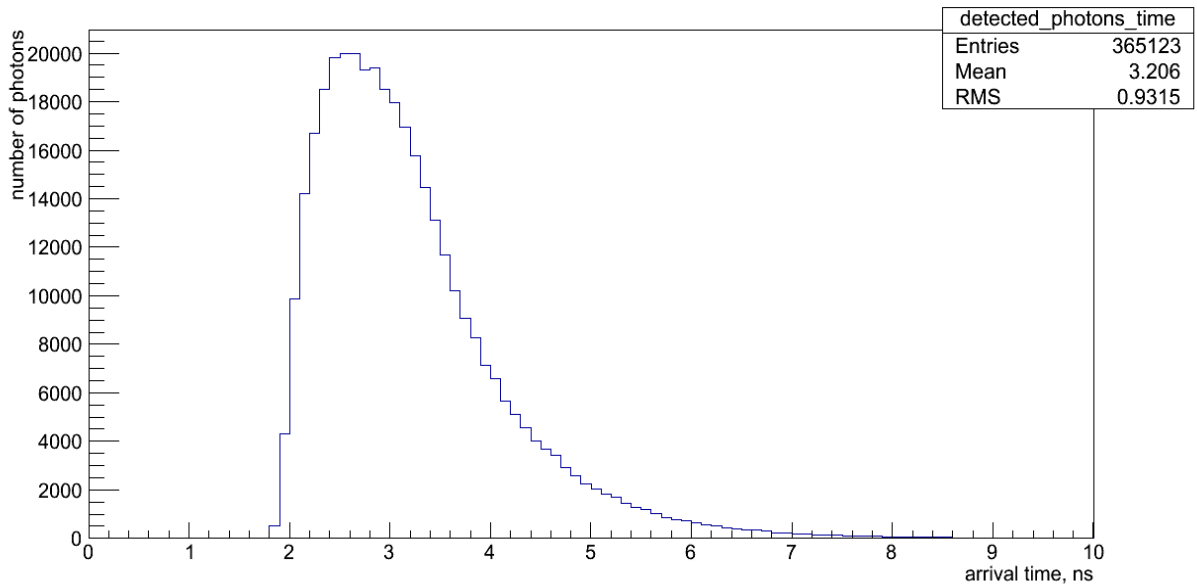

Figure 5: Detected photons arrival time distribution for GD.

## 4. Experimental measurements

### 4.1 MIP Signal Response Calibration

The response of each SD and GD to minimally ionizing particle (MIP) is calibrated in double coincidence setup using a secondary trigger comprised of MELTZ [12] FEU49 and a 15 cm diameter scintillator that is placed under each detector center during calibration process. This process yields the area of single MIP signal from each detector. In order to comply with analysis process, MIP area is defined between 10% and 90% of the pulse total area. MIP calibration for all detectors at different bias voltages is shown in [9].

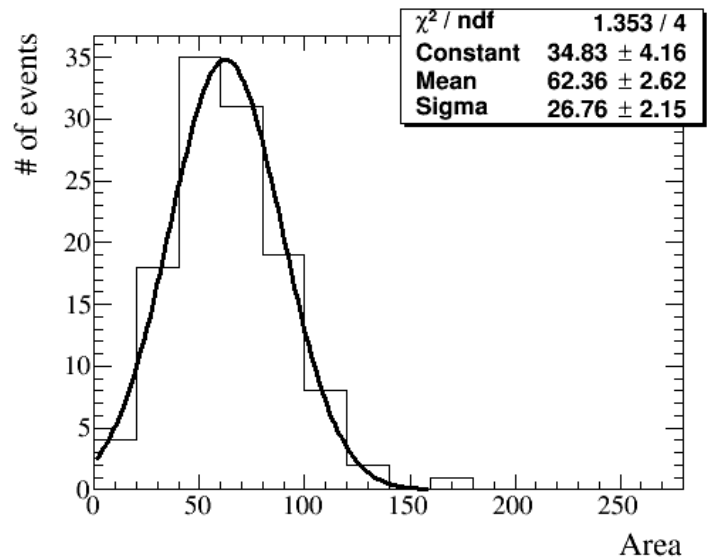

**Figure 6: R7723 PMT single PE pulse area at 1700V [9]**

In order to compare with the detector simulation, the single photo-electron (PE) area calibration for R7723 PMT is done in order to approximate photon count per MIP. Figure 6 shows the single PE pulse area at 1700V with pedestal subtracted. The single PE response is calibrated at different PMT bias voltages from 1300V to 2000V to cover biases of all detectors. The resultant single PE area vs bias voltage is shown in Figure 7**Error! Reference source not found.**.

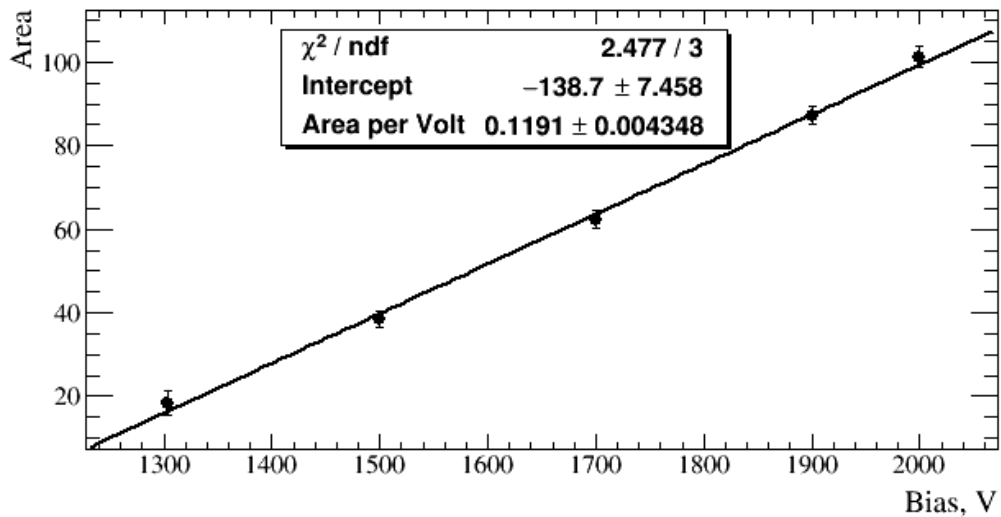

**Figure 7: Single PE pulse area vs. bias voltage for R7723 PMT [9]**

PE number for each scintillator and glass detector is calculated using MIP pulse area and single PE PMT response area of corresponding detector. Signal losses in each cable [13] and presence of impedance matching resistor are also taken into account. As a result, average PE number obtained for SD is $18.5 \pm 1.2$ photons, and PE number for GD is $3.6 \pm 0.4$ photons.

## 4.2 SD Detector Response Uniformity

SD detector has the pyramid-shaped enclosure with 100 cm x 100 cm x 5 cm plastic scintillator placed at the bottom, and a PMT placed 60 cm above the scintillator. The MIP signal calibration is done at the center of SD or GD, however, particles arrive randomly. Thus, a number of particles will be distributed uniformly across the detector area, and we need to measure the non-uniformity of detector response in order to estimate the charged particle flux through each SD.

In order to measure the non-uniformity, each SD is scanned using $^{60}$Co radioactive source across lines x = 20 cm, x = 80 cm and two diagonals. Output current from PMT with dark current (e.g. without rad. source) subtracted is recorded. Data is normalized to maximal value for uniformity comparison across different lines. The $I/I_{max}$ vs distance is shown in Figure 8.

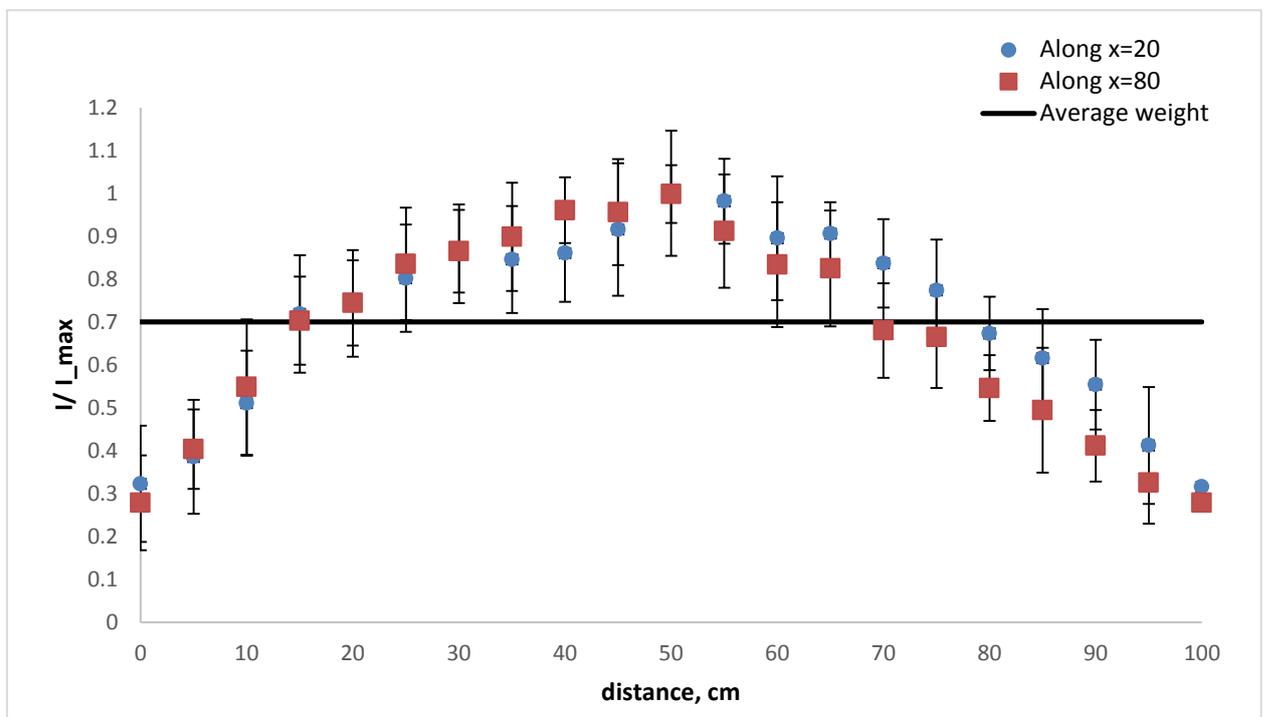

**Figure 8: SD Detector Response Uniformity.**

Light yield of scintillator itself is uniform across its volume; the rest is the effect of the detector shape. Based on that, the normalized detector response data is scaled to an average value by different weights. Average weight value is obtained by finding a line such as the area under it equals to the area under the data points. Average weight value for several SDs is $0.70 \pm 0.06$.

## 5. Conclusion

Design and implementation of GD are discussed. The main characteristics of GD obtained from simulation are presented, and simulation comparison to experimental measurements of SD characteristics is shown as a validation. According to the results, GD has a better uniformity than SD, but SD has higher light yield. The results are summarized in the Table

1 below. The simulation has been validated by scintillator with white edges options with close matching to measured values within corresponding uncertainties.

**Table 1: Summary of simulation results**

| Medium Type | Number of detected photons | Average signal width, ns | Uniformity coefficient |
|---|---|---|---|
| Glass, Black Edges | 3.65±0.03 | 1.8±0.1 | 0.74±0.03 |
| Glass, White Edges | 4.62±0.03 | 2.1±0.1 | 0.77±0.02 |
| Scint, Black Edges | 16.09±0.02 | 7.5±0.1 | 0.68±0.04 |
| Scint, White Edges | 18.92±0.02 | 9.2±0.1 | 0.70±0.04 |

# Bibliography

[1]  RU Beisembaev, EA Beisembaeva, OD Dalkarov, VA Ryabov, AV Stepanov, NG Vildanov, MI Vildanova, VV Zhukov, KA Baigarin, D Beznosko, TX Sadykov, NS Suleymenov, "The 'Horizon-T' Experiment: Extensive Air Showers Detection," arXiv:1605.05179 [physics.ins-det], May 17 2016.

[2]  Adil Baitenov, Alexander Iakovlev, Dmitriy Beznosko, "Technical manual: a survey of scintillating medium for high-energy particle detection," *arXiv:1601.00086,* 2016/1/1.

[3]  Hamamatsu Corporation, 314-5,Shimokanzo, Toyooka-village, Iwatagun,Shizuoka-ken, 438-0193 Japan. http://www.hamamatsu.com.

[4]  JSC 'LZOS' 140080, Lytkarino, Moscow region, Russia, H.1, Parkovay Str.http://www.lzos.ru/.

[5]  "Cherenkov Radiation," Emma Ona Wilhelmi, 18 October 2001. [Online]. Available: http://www.gae.ucm.es/~emma/docs/tesina/node4.html.

[6]  "Modeling Diffuse Reflection (or How to Sample Cosine Distribution)," 2 April 2015. [Online]. Available: https://www.particleincell.com/2015/cosine-distribution/.

[7]  R. Brun, F. Rademakers, "ROOT: An object oriented data analysis framework," *Nucl. Instrum. Meth. A,* vol. 389, p. 81–86, 1997.

[8]  Duspayev, A., R. U. Beisembaev, T. Beremkulov, D. Beznosko, A. Iakovlev, K. Yelshibekov, M. Yessenov, V. Zhukov, "Simulation, design and testing of the HT-KZ Ultra-high energy cosmic rays detector system," *Proceedings of ICHEP2016,* vol. PoS(ICHEP2016)721, 2016.

[9]  D. Beznosko, T. Beremkulov, A. Iakovlev, A. Duspayev, M. I. Vildanova, T. Uakhitov, K. Yelshibekov, M. Yessenov, V.V. Zhukov, "Horizon-T Experiment Calibrations – MIP Signal from Scintillator and Glass Detectors," *arxiv:1703.07559,* 3/2017.

[10] I.M., Sobol', The Monte-Carlo methods, Moscow: "Nauka", 1973.

[11] G. Japaridze, M. Ribordy, "Realistic arrival time distribution," *arXiv:astro-ph/0506136v1,* 2005.

[12] MELZ-FEU, 4922-y pr-d, 4c5, Zelenograd, g. Moskva, Russia, 124482 (http://www.melz-feu.ru).

[13] D Beznosko, T Beremkulov, A Iakovlev, Z Makhataeva, M I Vildanova, K Yelshibekov, V V Zhukov, "Horizon-T Experiment Calibrations-Cables," *arXiv:1608.04312,* 8/2016.

[14] Duspayev et al., "The distributed particle detectors and data acquisition modules for Extensive Air Shower measurements at "Horizon-T KZ" experiment," in *PoS(PhotoDet2015)056, in proceedings to PhotoDet2015 conference*, Moscow, 2015.

[15] S Assylbekov et al., "The T2K ND280 off-axis pi–zero detector," *Nuclear Instruments and*

*Methods in Physics Research Section A,* vol. 686, pp. 48-63, 2012/9/11.

[16] A. Dyshkant, D. Beznosko, G. Blazey, E. Fisk, E. Hahn, V. Rykalin, M. Wayne and V. Zutshi, "SCINTILLATION DETECTORS-Quality Control Studies of Wavelength Shifting Fibers for a Scintillator-Based Tail Catcher Muon Tracker for Linear Collider Prototype Detector," IEEE Transactions on Nuclear Science, volume 53, issue 6, page 3944, 2006.

[17] Lindsey J Bignell at al., "Characterization and Modeling of a Water-based Liquid Scintillator," arXiv:1508.07029, Aug 2015.

[18] D. Beznosko, G. Blazey, A. Dyshkant, V. Rykalin, V. Zutshi, "Effects of the Strong Magnetic Field on LED, Extruded Scintillator and MRS Photodiode," *NIM A,* vol. 553, pp. 438-447, 2005.

[19] D. Beznosko, T. Beremkulov, A. Duspayev, A. Iakovlev, "Random Number Hardware Generator Using Geiger-Mode Avalanche Photo Detector," in *PoS(PhotoDet2015)049, in proceedings to PhotoDet2015*, Moscow, 2015.

[20] D. Beznosko, T. Beremkulov, A. Duspayev, A. Tailakov and M. Yessenov, "Random Number Hardware Generator Using Geiger-Mode Avalanche Photo Detector," arXiv:1501.05521 [physics.ins-det], 2015/1/22.

[21] D. Beznosko, A. Dyshkant, C.K. Jung, C. McGrew, A. Pla-Dalmau, V. Rykalin, ""MRS Photodiode Coupling with Extruded Scintillator via Y7 and Y11 WLS Fibers"," February FERMILAB-FN-0796, 2007.

[22] L. J. Bignell at al., "Characterization and Modeling of a Water-based Liquid Scintillator," *Journal of Instrumentation, IOP Publishing,* vol. 10, p. 12009, 12/2015.